\documentclass[12pt]{article}
\usepackage{epsfig,amssymb,mathrsfs}

\def\beq{\begin{equation}}
\def\eeq#1{\label{#1}\end{equation}}
\def\eeqn{\end{equation}}

\def\beqa{\begin{eqnarray}}
\def\eeqa#1{\label{#1}\end{eqnarray}}
\def\eeqan{\end{eqnarray}}

\let\bar=\overbar

\def\Dslash{\not{\hbox{\kern-4pt $D$}}}
\def\dslash{\not{\hbox{\kern-2pt $\del$}}}

\def\msb{{\bar{\ssstyle M \kern -1pt S}}}

\usepackage{fancyhdr,graphicx}
\def\Title#1{\begin{center} {\Large {\bf #1} } \end{center}}

\begin{document}

\Title{Dynamical chiral symmetry breaking in strangelets at finite
temperature}

\bigskip\bigskip


\begin{raggedright}

{\it Chang-Qun Ma\index{Ma, C. Q} and
Chun-Yuan Gao\index{Gao, C. Y.}\\
School of Physics and State
Key Laboratory of Nuclear Physics and Technology\\
Peking University\\
Beijing 100871\\
P. R. China\\
{\tt Email: gaocy@pku.edu.cn}}
\bigskip\bigskip
\end{raggedright}

\section{Introduction}\label{intro}\noindent
Strangelet\cite{strangelet} has been a hot topic since it was
conjectured that strange quark matter might be the absolutely stable
state \cite{supposesqm1,supposesqm2}. It is expected that
strangelets with low masses might be detected in heavy ion
collisions\cite{Barrette}, where the ejected particles are formed in
hot environment. Searching for strangelets in terrestrial
laboratories is important not only for understanding the strong
interaction physics, but also because that strangelets are regarded
as the unambiguously detected signals for Quark Gluon Plasma(QGP)
production\cite{qgpsignal,qgpsignal2}. So it is valuable to study
the properties of strangelets at finite temperature.

Strange quark matter may not be absolutely stable when realistic
current quark masses are introduced which has its origin in the
spontaneous breaking of chiral symmetry\cite{bo}. Chiral symmetry
restoration is enhanced by the finite size effects for droplets with
relatively small baryon numbers\cite{kh}. Therefore the chiral
symmetry breaking is expected to have great influences on the
stability and bulk properties of strangelets. In the present paper,
the chiral symmetry breaking in strangelets at finite temperature is
investigated.

Strangelets at both zero and finite temperature have been frequently
studied in the framework of MIT bag model, and many extraordinarily
important results were
presented\cite{mit_fint1,mit_sta,mit_fint2,mid_madsen,mit_shell,mid_madsen2}
in the last decades. However, the essential supposition of the MIT
bag model is that quarks are dealt with as free Fermi gas and
thereafter their masses take their current masses\cite{mitmodel}.
Therefore, the dynamical chiral symmetry breaking is absent in the
MIT bag model. In Refs.\cite{chiral1,chiral2,chiral3}, the NJL model
is adopted to study the dynamical chiral symmetry breaking in the
strangelets at zero temperature, and it is found that the chiral
symmetry would break spontaneously at some critical radius. For
strangelets, finite size effect should be considered and the
multiple reflection expansion (MRE) approximation\cite{mre} helps to
settle this problem well for strangelets with large baryon numbers.
While the shell effect is vitally important for strangelets with
small baryon numbers and the MRE approximation would lead to
considerable error\cite{mit_shell}. Recently, Yasui brought forward
a NJL$+$MIT bag model\cite{njlmit,njlmit2}, in which the NJL model
was adopted to describe quark fields inside strangelets and
calculations involving mode filling in spherical MIT bag were
carried out. The NJL$+$MIT bag model respects the shell effect and
quark masses in the model are dynamically generated. So the model is
suitable for studying the dynamical chiral symmetry breaking in
strangelets with small baryon numbers at zero temperature.

In the present work, we extend the NJL$+$MIT bag model in quantum
statistical approach to study the strangelets at finite temperature.
We construct a grand canonical partition function using the discrete
eigenenergies of quarks and the energy contributions from vacuums
are taken into account. In addition, the t'Hooft term, which models
the U(1)$_{A}$ symmetry breaking of QCD and has been neglected in
Yasui's model, is also considered in the present NJL Lagrangian. In
the present model, we focus on the strangelets at $\beta$
equilibrium only, that means the chemical potentials for quarks with
three flavors are equal because the electrons would not exist in
strangelets (because the de Broglie wave length of the electron is
larger than the size of strangelets\cite{chiral2}). However, it will
be straightforward to extend our model to describe strangelets being
not in $\beta$ equilibrium, which are likely to be produced in heavy
ion collisions.

\section{Model}\label{model}\noindent
The strangelet is described by a static spherical MIT bag with
radius $R$, in which valence quarks are confined. The NJL Lagrangian
with t'Hooft term\cite{njl} is used to describe the quark fields
inside the bag. As in the MIT bag model, a surface interacting term
$\mathcal {L}_\delta$ is introduced to ensure that the vector quark
current is continuous on the surface\cite{mitboundary}. With these
assumptions, the total Lagrangian density is:
\begin{eqnarray}\label{lg}
{\cal L}_{QD}=\left[\bar{q}\left({\rm
i}\gamma^\mu\partial_\mu-\hat{m}_0\right)q+{\cal
L}_{\bar{q}q}\right]\theta(R-r)+{\cal L}_{\delta},
\end{eqnarray}
where
\begin{eqnarray}\label{chiral_term}
{\cal L}_{\bar{q}q}=\!\!\!\!\!&&
G\sum\limits_{a=0}^{8}\left[\left(\bar{q}\tau_{a}q\right)^2+\left(\bar{q}{\rm
i}\gamma_5\tau_aq\right)^2\right]\nonumber
\\
&&-K\left\{{\rm{det}}_{f}\left[\bar{q}(1+\gamma_5)q\right]+{\rm{det}}_f
\left[\bar{q}(1-\gamma_5)q\right]\right\}\end{eqnarray} and
\begin{eqnarray}
{\cal L}_{\delta}=-\frac{1}{2}\bar{q}q\delta(r-R).
\end{eqnarray}
And $\hat{m}_0={\rm{diag}}(m_{0u},m_{0d},m_{0s})$ is the current
quark mass matrix in flavor space; the step function $\theta$ is to
confine the quarks inside the MIT bag. It can be seen that when
$R\rightarrow$$+\infty$, the Lagrangian of (\ref{lg}) reduces to the
form that describes quark matter in bulk.

The surface interacting term $\mathcal {L}_\delta$ breaks the chiral
symmetry explicitly, and in principle a pion cloud coupling with the
bag surface could be introduced to recover the broken chiral
symmetry. However, it is found that the pion cloud contribution is
negligible for strangelets with baryon number
A$\gtrsim$5\cite{njlmit2}. In the following discussions, we will
focus on strangelets with baryon numbers of the order of 100, so the
present treatment of the surface interacting term is justified.

In the present work, we restrict ourselves to mean-field
approximation and focus on the chiral condensates defined as
\begin{equation}
\phi_f=\left<\bar{q}_fq_f\right>,\hskip5mm f=u,d,s.
\end{equation}
After bosonization, one obtains the linearized version of the model
in the mean-field approximation,
\begin{eqnarray}\label{meanl}
{\cal L}_{QD}=\!\!\!\!\!\!\!&&\left[\bar{q}\left({\rm
i}\gamma^\mu\partial_\mu-\hat{m}\right)q-2G\sum\limits_f\phi_f^2
+4K\phi_u\phi_d\phi_s\right]\nonumber\\
&&\times\theta(R-r)-\frac{1}{2}\bar{q}q\delta(r-R),
\end{eqnarray}
where the constituent quark mass matrix is
\begin{equation}
\hat{m}=\left(
                    \begin{array}{ccc}
                      m_u&0&0\\
                      0&m_d&0\\
                      0&0&m_s
                    \end{array}
                  \right),
\end{equation}in which
\begin{eqnarray}\displaystyle
&&m_i=m_{0i}-4G\phi_i+2K\phi_j\phi_k,\\\nonumber &&(i, j, k) = {\rm
any\ permutation\ of\ } (u, d, s).\end{eqnarray}

Then the field equation for $r<R$ and boundary condition at $r=R$
can be derived from (\ref{meanl}) by variational principle, they
are:
\begin{eqnarray}\label{eigenequ}
\left\{ \begin{array}{l}
         \left({\rm i}\gamma^\mu\partial_\mu-\hat{m}\right)q=0,~~~~~r<R, \\
         -{\rm i}\vec{n}\cdot\vec{\gamma}q=q,~~~~~~~~~~~~~r=R.
       \end{array} \right.
\end{eqnarray}
Solve the Eq.(\ref{eigenequ}) in spherical coordinates, the
eigenstates for quarks with flavor $f$ are obtained
\begin{eqnarray}\label{eigenfun}
q_{f}(\vec{r})=\mathcal {N}\left(\begin{array}{c}
                                                    j_l(pr) \\
                                                    \displaystyle\epsilon\frac{{\rm i}p}{E+m_f}j_{l+\epsilon}(pr)\vec{\sigma}\cdot\vec{n}
                                                  \end{array}\right)\mathcal
                                                  {Y}_{jm}^l(\vec{n}).
\end{eqnarray}
The meanings of quantum numbers are: the orbital angular momentum
$l$; the total angular momentum $j=l+\epsilon/2$($\epsilon=\pm1$);
the third component projection $m$ of the total angular momentum and
the eigenmomentum $p$. For convenience, we'll use $\alpha$ to
represent the quantum number set $(l,j,m)$ in the following. And
$\vec{n}\equiv\vec{r}/|\vec{r}|$, $\mathcal {N}$ is the
normalization constant, $j_l$ is the spherical Bessel function of
rank $l$, $\mathcal {Y}_{jm}^l$ is the eigenstates of the total
angular momentum, $E\equiv\sqrt{p^2+m_f^2}$.

The eigenstates (\ref{eigenfun}) are for both positive energy states
and negative energy states. In positive energy states, eigenmomentum
$p$ should satisfy the boundary condition
\begin{eqnarray}\label{posmom}
j_l(pR)=\epsilon\frac{p}{E+m_f}j_{l+\epsilon}(pR),
\end{eqnarray}
while the boundary condition in negative energy states is
\begin{eqnarray}\label{negmom}
j_{l+\epsilon}(pR)=-\epsilon\frac{p}{E+m_f}j_l(pR).
\end{eqnarray}
By introducing the node quantum number $n$, we can then denote the
solutions for (\ref{posmom}) and (\ref{negmom}) as $p_{f,\alpha}^n$
and $\overline{p}_{f,\alpha}^n$ respectively.

Then, the energies for quarks with flavor $f$ in positive energy
states $(f,\alpha,n)$ and negative energy states $(f,\alpha,n)$ are:
\begin{eqnarray}\label{potnegeng}
E_{f,\alpha}(p_{f,\alpha}^n)=\sqrt{{{p_{f,\alpha}^n}}^2+m_f^2},\\
\overline{E}_{f,\alpha}(\bar{p}_{f,\alpha}^n)=-\sqrt{{{\overline{p}_{f,\alpha}^n}}^2+m_f^2}.
\end{eqnarray}
In addition, the energy for hole states (antiquarks)
$(\bar{f},\bar{\alpha},\bar{n})$ which would be excited at finite
temperature is:
\begin{eqnarray}\label{holeeng}
E_{\bar{f},\bar{\alpha}}(p_{\bar{f},\bar{\alpha}}^{\bar{n}})=-\overline{E}_{f,\alpha}(\bar{p}_{f,\alpha}^n)=\sqrt{{{\overline{p}_{f,\alpha}^n}}^2+m_f^2}.
\end{eqnarray}

Now, we have solved out the single particle (SP) states for quarks
inside the bag. In the following, we'll construct the partition
function for quarks and antiquarks in terms of SP energies within
the particle number representation.

The particle number operators in positive energy states and hole
states are denoted as $\hat{N}_{f,\alpha,n}$ and
$\hat{N}_{\bar{f},\bar{\alpha},\bar{n}}$, respectively. The
Hamiltonian for a strangelet is
\begin{eqnarray}\label{har}
\hat{H}=\!\!\!\!\!\!\!\!&&\sum_{f,\alpha,n}{\nu}\hat{N}_{f,\alpha,n}E_{f,\alpha}(p_{f,\alpha}^n)+\sum_{\bar{f},\bar{\alpha},\bar{n}}{\nu}\hat{N}_{\bar{f},\bar{\alpha},\bar{n}}E_{\bar{f},\bar{\alpha}}(p_{\bar{f},\bar{\alpha}}^{\bar{n}})\nonumber\\
&&+E_{\rm sea}+V_{\rm mean}-E_{\rm vac},
\end{eqnarray}
where $\nu=3$ is the degenerate degree in color space; $E_{\rm sea}$
is the energy of the confined vacuum fulfilled by sea quarks in the
bag, which is
\begin{eqnarray}\label{esea}
E_{\rm sea}=\sum_{f,\alpha,n}^{\rm
sea}{\nu}\overline{E}_{f,\alpha}(\overline{p}_{f,\alpha}^n);
\end{eqnarray}
$V_{\rm mean}$ is the potential energy brought by the mean fields
and its form takes
\begin{eqnarray}\label{vmean}
V_{\rm
mean}=\frac{4}{3}{\pi}R^3\left(2G\sum\limits_f\phi_f^2-4K\phi_u\phi_d\phi_s\right).
\end{eqnarray}
We have subtracted $E_{\rm vac}$(=$4{\pi}R^3\varepsilon_{\rm
vac}/3$) from the Hamiltonian because the measured energy density is
the difference between the total energy density of the strangelet
and the energy density of the vacuum where the strangelet is absent.
$E_{\rm vac}$ is obtained by the NJL model for quark matter in bulk
(taking the limit $R\rightarrow$$+\infty$ in Eq.(\ref{lg})).

Then the partition function can be constructed as
\begin{eqnarray}\label{pf}
\Xi\!\!\!&=&\!\!\!\mathrm{Tr}\exp\left[-\beta(\hat{H}-{\mu}\hat{N})\right]\nonumber\\
\!\!\!&=&\!\!\!\mathrm{Tr}\exp\left\{-\beta\left[\hat{H}-{\mu}\left(\sum_{f,\alpha,n}\nu\hat{N}_{f,\alpha,n}-\sum_{\bar{f},\bar{\alpha},\bar{n}}\nu\hat{N}_{\bar{f},\bar{\alpha},\bar{n}}\right)\right]\right\}\nonumber\\
\!\!\!&=&\!\!\!\exp\left[-\beta\left(E_{\rm sea}+V_{\rm mean}-E_{\rm vac}\right)\right]\nonumber\\
\!\!\!&&\!\!\!\times\prod_{f,\alpha,n}\left\{1+\exp\left[-\beta\left(E_{f,\alpha}(p_{f,\alpha}^n)-\mu\right)\right]\right\}^{\nu}\nonumber\\
\!\!\!&&\!\!\!\times\prod_{\bar{f},\bar{\alpha},\bar{n}}\left\{1+\exp\left[-\beta\left(E_{\bar{f},\bar{\alpha}}(p_{\bar{f},
\bar{\alpha}}^{\bar{n}})+\mu\right)\right]\right\}^{\nu},
\end{eqnarray}
and
\begin{eqnarray}\label{lgpf}
\ln\Xi\!\!\!&=&\!\!\!-\beta\left(E_{\rm sea}+V_{\rm mean}-E_{\rm vac}\right)\nonumber\\
\!\!\!&&\!\!\!+\sum_{f,\alpha,n}\nu\left\{1+\exp\left[-\beta\left(E_{f,\alpha}(p_{f,\alpha}^n)-\mu\right)\right]\right\}\nonumber\\
\!\!\!&&\!\!\!+\sum_{\bar{f},\bar{\alpha},\bar{n}}\nu\left\{1+\exp\left[-\beta\left(E_{\bar{f},\bar{\alpha}}(p_{\bar{f},\bar{\alpha}}^{\bar{n}})+\mu\right)\right]\right\},\nonumber\\
\end{eqnarray}
where $\beta=1/T$ and $T$ is the temperature.

So the free energy of a strangelet is:
\begin{eqnarray}\label{freeenergy}
F=-\frac{1}{\beta}\ln\Xi+{\mu}N.
\end{eqnarray}
where $N$ is the total quark number, which satisfies
\begin{eqnarray}\label{quarknum}
N=\frac{1}{\beta}\frac{\partial}{\partial{\mu}}\ln\Xi.
\end{eqnarray}
For thermodynamical stable states, the free energy should take its
minimum. So the order parameters $\phi_f$ of dynamical chiral
symmetry breaking should be solved by minimizing the free energy,
i.e.
\begin{eqnarray}\label{phi}
\frac{{\partial}F}{{\partial}\phi_f}=0,~~~~~~f=(u,d,s).
\end{eqnarray}
For stable strangelets in vacuum the pressure should be zero. This
condition is used to determine the bag radius and it means
\begin{eqnarray}\label{pressure}
{\cal P}=-\left[\frac{\partial}{\partial{V}}F(T,R)\right]_{T,N}=0,
\end{eqnarray}
where $V=4\pi{R^3}/3$ is the volume of a strangelet.

The NJL model is nonrenormalizable, so a cutoff in momentum space is
often used to regularize the infinite summation of SP states. In the
present model, we take the form as that in Ref.\cite{njlmit}
\begin{eqnarray}\label{cutoff}
g(p/\Lambda)=\frac{1}{1+(p/\Lambda)^\lambda}.
\end{eqnarray}
Then in numerical calculations, the following substitutions should
be made:
\begin{eqnarray}
\sum_{f,\alpha,n}&\longrightarrow&\sum_{f,\alpha,n}g(p_{f,\alpha}^n/\Lambda),\nonumber\\
\sum_{f,\alpha,n}^{\rm sea}&\longrightarrow&\sum_{f,\alpha,n}^{\rm sea}g(\bar{p}_{f,\alpha}^n/\Lambda),\nonumber\\
\sum_{\bar{f},\bar{\alpha},\bar{n}}&\longrightarrow&\sum_{\bar{f},\bar{\alpha},\bar{n}}g(p_{\bar{f},\bar{\alpha}}^{\bar{n}}/\Lambda).\nonumber
\end{eqnarray}

There are 6 parameters in our model totally. Five of them, such as
the current quark masses $m_{0s}$ and $m_{0q}$, couplings $G$ and
$K$ and the cutoff $\Lambda$, have been determined in
Ref.\cite{ours} previously. They are: $m_{0q}$=0.0055GeV,
$m_{0s}$=0.1409GeV, $\Lambda$=0.6028GeV, $G$$\Lambda^{2}$=1.803,
$K$$\Lambda^{5}$=12.93. This parameter set reproduces the meson
masses $m_\pi=134.98$MeV, $m_K=497.65$MeV and $m_{\eta'}=957.78$MeV
and the $\pi$ decay constant $f_{\pi}=92.2$MeV. We follow the method
in Ref.\cite{njlmit} to adjust the diffuseness parameter $\lambda$
in Eq.(\ref{cutoff}) to fit the baryons' mass,
$(m_N+m_\Delta)/2=1.1$GeV, in vacuum. We get $\lambda=27$. Finally,
the value of the energy density of the vacuum in bulk
$\varepsilon_{\rm vac}$ is fixed once the model parameters are
determined, and here $\varepsilon_{\rm vac}=-4.416$GeV$/$fm$^3$.

\section{Results and Discussions}
We focus on the dynamical chiral symmetry breaking, which means that
the chiral symmetry breaks spontaneously, at finite temperature. We
find that the spontaneous chiral symmetry breaking is restored
inside the strangelets with baryon number $A\lesssim$150 when
$T\lesssim$100MeV, so the quark masses for such strangelets take
their current mass values. We take the strangelets with total baryon
numbers $A=$240, 300 and 400 as numerical examples here.

\begin{figure*}[t]\centering
{\epsfig{file=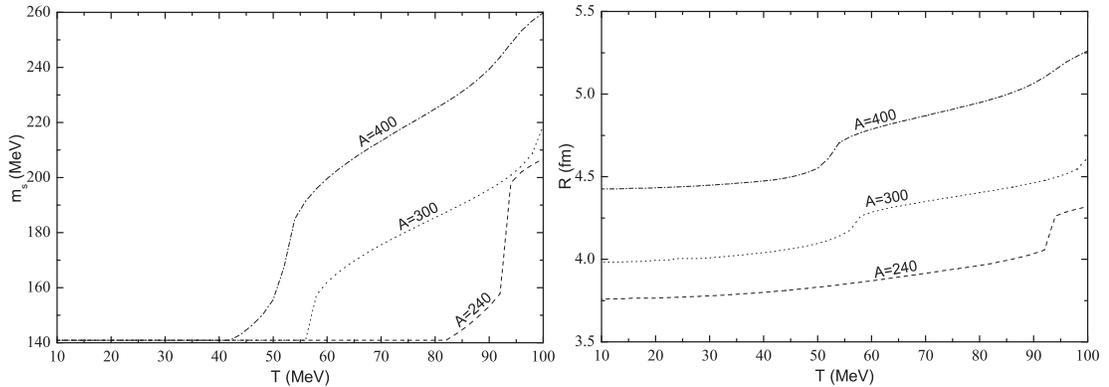,width=14.5cm}} \caption{The left panel
is the mass of strange quark as a function of temperature; the right
panel gives the relation between radii of strangelets with fixed
masses and temperature. The three curves in both panels are for
strangelets with baryon numbers of 240, 300 and 400,
respectively.}\label{fig1}\end{figure*}

The strange quark masses within strangelets are plotted in the left
panel of FIG. \ref{fig1}. Since the condensations $\phi_u$ and
$\phi_d$ vanish when $T\le$100MeV, the effective masses of
nonstrange quarks take their current masses and then we are not
going to plot them here. We can see that the chiral symmetries break
spontaneously inside the strangelets with $A=$240, 300 and 400 at
critical temperatures $T\backsimeq$83MeV, 57MeV and 43MeV
respectively, then the strange quarks start to gain their masses by
self-energy. And the effective masses of strange quarks begin to
increase as the temperature rises, which means that the chiral
symmetry will break to a larger extent as the temperature rises. And
from the three curves of different baryon numbers, we can see that
the chiral symmetries breaks to a larger extent, which is reflected
by a larger strange quark mass at a given temperature, inside the
strangelets with larger baryon numbers.

\begin{figure}[h]\centering
{\epsfig{file=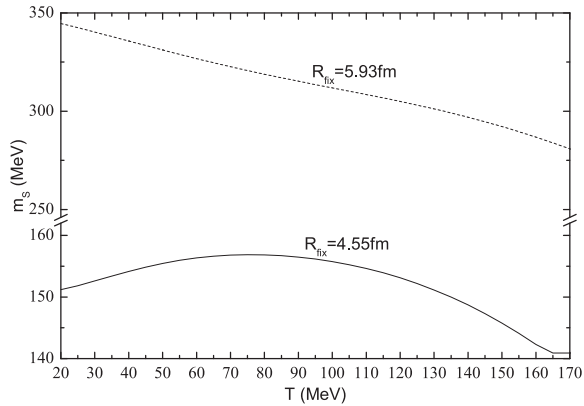,width=7.6cm}} \caption{The effective strange
quark mass as a function of temperature with the different fixed
radius of the bag. The baryon number of the strangelet is taken as
400.}\label{fig2}\end{figure}

Whereas in the standard conjectured QCD phase diagram\cite{alford},
the spontaneous breaking of chiral symmetry inside the nuclear
matter in bulk tends to restore as temperature increases, where
there is no external constraints imposed (like the pressure) and the
finite size effect is absent. Here we have seen that the strange
quark mass relating to the spontaneous breaking of chiral symmetry
becomes larger as the temperature increases for the strangelets with
zero pressure. It seems abnormal.

From the radii of strangelets in the right panel of FIG. \ref{fig1},
we can see that the volume of a strangelet is expanding as the
temperature increases to keep the pressure at zero. Especially,
there is a steep rise of the radius near the critical temperature.
Because that the baryon number is conserved, the volume expanding
leads to the decreasing of density inside the strangelets. In the
standard conjectured QCD phase diagram, the chiral symmetry tends to
spontaneously break to a larger extent as the density decrease.
Therefore, we conclude that one reason for the abnormal phenomenon
is the volume expanding of the strangelets. To corroborate this, we
fix the radius of the strangelet artificially at 4.55fm (it is the
stable radius at temperature 50MeV) for the strangelet with $A=$400
and the results are plotted with a solid line in FIG. \ref{fig2}. We
can see that the strange quark mass decreases when $T\gtrsim$75MeV,
and it will eventually decrease to its current mass value at higher
temperature. Therefore, our argument is approved.

But we can see that the strange quark mass still increases slowly as
the temperature rises for low temperatures (when $T\lesssim$75MeV).
This reveals that the dynamical chiral symmetry breaking is also
affected by the finite size effect of the strangelet, which is
usually explained as surface energy and curvature energy in the MRE
approximation. In order to illustrate the second reason, we increase
the fixed radius of the strangelet. Because the Lagrangian density
of strangelet (${\cal L}_{QD}$ in Eq.(\ref{lg})) will gradually
reduce to the form that describes quark matter in bulk as the radius
$R$ increases, it is imaginable that the discrete distribution of
eigenstates in momentum space will gradually reduce to the
continuous distribution as in bulk quark matter when the radius
increases, then the finite size effect will become less important
correspondingly. The mass of strange quark inside the strangelet
with a larger fixed radius 5.93fm (it is 30GeV$^{-1}$ in natural
unit) are plotted in FIG. \ref{fig2} with a dashed line. It can be
seen that the abnormal phenomenon that the strange quark mass
increases as the temperature rises vanishes now. And it is also
expected that the strange quark mass will eventually decrease to its
current mass value at sufficiently high temperature in this case. In
conclusion, the strange quark mass inside the strangelets increases
as the temperature rises is caused by two reasons: 1) the volume
expanding and 2) the finite size effect.

\begin{figure}[h]\centering
{\epsfig{file=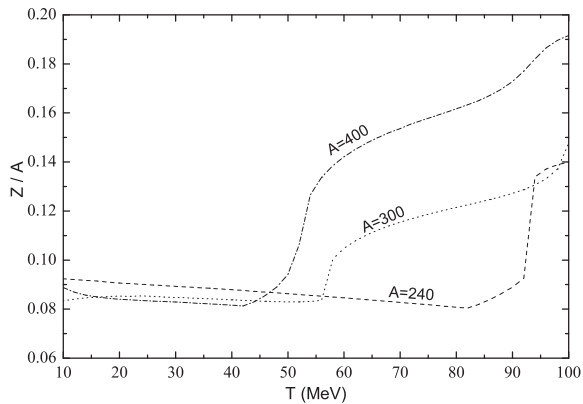,width=7.6cm}} \caption{Charge-mass
ratios as functions of temperature for strangelets with different
masses.}\label{fig3}\end{figure}

The chiral symmetry breaking would affect the bulk properties of
strangelets, one of which is the charge-mass ratio $Z/A$, i.e. the
proportion between the carried charge and the total baryon number of
a strangelet. The value of charge-mass ratio is one of the most
import physical quantity for distinguishing strangelets in
experiments\cite{searchsqm}. The charge-mass ratios as functions of
temperature are plotted in FIG. \ref{fig3}. We can see that before
the spontaneous chiral symmetry breaking happens, the charge-mass
ratio is $\sim$ 0.1, which is the same as in MIT bag model. While
after the chiral symmetry breaks spontaneously, the charge-mass
ratio begins to increase and it reaches the value of $\sim$0.2 at
the temperature 100MeV for the strangelet with $A=$400.
\section{Summary and conclusions}\label{sec4}\noindent
Considering a NJL type interaction for quarks confined in a MIT bag,
we build a model to describe the strangelets at finite temperature,
in which quark masses are dynamically generated. We study the
dynamical chiral symmetry breaking which relates to the effective
quark masses inside strangelets at finite temperature. It is found
that when temperature is under 100MeV, there is no spontaneous
breaking of chiral symmetry inside the strangelets with baryon
numbers $A<$150. By taking the strangelets with baryon numbers of
240, 300 and 400 as numerical examples, we find that: 1) the chiral
symmetry breaks spontaneously at some finite temperatures and the
strange quark mass will increase as temperature rises; 2) the
dynamical chiral symmetry breaking leads to a considerable change of
the charge-mass ratio of the strangelets. We illustrate the
``abnormal" phenomenon that the strange quark mass inside the
strangelets increases as temperature rises by two reasons: 1) the
volume expanding and 2) the finite size effect.
\bigskip
Financial support by the National Natural Science Foundation of
China under grants 10875003, 10811240152, 10475002 \& 10435080 is
gratefully acknowledged.

\end{document}